\documentclass[12pt]{iopart}
\usepackage[dvips]{graphicx}
\usepackage{units}
\usepackage{ae}

\begin{document}
%
%-----------title---------------
\title{Temperature independent current deficit due to induced quantum nanowire vibrations}
\author{Gustav Sonne}
\address{Department of Physics, University of Gothenburg, SE-412 96 G\"oteborg, Sweden}
\ead{gustav.sonne@physics.gu.se}
\date{\today}
%
%----abstract
\begin{abstract}
We consider electronic transport through a suspended voltage-biased
nanowire subject to an external magnetic field. In this paper, we show
that the transverse magnetic field, which acts to induce coupling
between the tunnelling current and the vibrational modes of the wire,
controls the current-voltage characteristics of the system in novel
ways. In particular, we derive the quantum master equation for the
reduced density matrix describing the nanowire vibrations. From this
we find a temperature- and bias voltage-independent current deficit in
the limit of high bias voltage since the current through the device is
lower than its value at zero magnetic field. We also find that the
corrections to the current from the back-action of the vibrating wire
decay exponentially in the limit of high voltage. Furthermore, it is
shown that the expression for the temperature- and bias
voltage-independent current deficit holds even if the nanowire
vibrational modes have been driven out of thermal equilibrium.
\end{abstract}
%73.23.-b Electronic transport in mesoscopic systems
%73.40.GK Tunneling
%73.63.Fg Nanotubes
\submitto{\NJP} 
\tableofcontents
\maketitle
\section{Introduction}
Nanoelectromechanical systems (NEMS) are
mesoscopic devices whose functionality depends on the possibility to
induce mechanical vibrations or displacements of one or several of
their components \cite{Craighead}. Examples of such setups are
numerous and include shuttling of single electrons and Cooper pairs
\cite{Shekhter2003,Shekhter2007,Erbe2001,Scheible2004,Moskalenko2008}, tuning of
mechanical bending vibrations of suspended
nanowires \cite{Jonsson2007,Jonsson2008,Sazonova} and mechanically
mediated superconducting and magnetic proximity effects
\cite{Isacsson2002,Fedorets2005,Gorelik2003} to name but a few.
Crucial advantages of the downscaling implied by the acronym NEMS are
the high (RF) vibration frequencies and the unprecedented sensitivity
to external stimuli that can be achieved. This is due in turn to the
low masses of these systems and to the strong coupling between
mechanical and electrical degrees of freedom at the nanometer length
scale, see, e.g., \cite{Lassagne2008,Jensen2008,Blencowenano} and
references therein. Also, nanoelectromechanical systems border the
world of quantum mechanics, which opens up the possibility to
experimentally study quantum effects on the interaction between
electrical and mechanical degrees of freedom in mesoscopic
systems \cite{LaHaye,Naik,Leroy,Knobel,Schwab2005}.

In this paper we will consider the nanoelectromechanical system
studied by Shekhter \etal \cite{Shekhter}, who analyzed the linear
conductance through a suspended voltage biased single-walled carbon
nanotube (SWNT) in the presence of a magnetic field. The main result
of \cite{Shekhter} was a prediction by the authors of a finite
negative magnetoconductance at low temperatures. This result is due to
a magnetic-field induced coupling of the electrons and the quantum
nanomechanical degrees of freedom in the system, which leads to an
effective multiconnectivity of the vibrating nanotube. More precisely,
the predicted result was attributed to two effects. One is the
suppression of the probability for electrons to tunnel through the
nanotube in the elastic channel, where the suppression is caused by
destructive quantum interference effects among multiple electron
tunneling paths. The other effect is Pauli-principle restrictions on
the inelastic tunneling channels. The predicted result is a
low-temperature effect, because at high temperatures, where the
Pauli-principle restrictions are lifted, the reduction of the
probability for tunneling through the elastic channel is fully
compensated by an increased probability for inelastic
tunneling. Hence, in this case, the linear conductance exactly
coincides with the transmission through the non-vibrating wire (see
also discussion in \cite{Shekhter}).

Here we will consider the same system as in \cite{Shekhter}, but now
beyond the linear bias-voltage regime. In particular we will show that
at large enough voltages there is a current deficit, as compared to
the non-vibrating wire. Furthermore, this current deficit is shown to
be independent of both the temperature and the bias voltage, making
this system a good candidate device for detection of quantum
vibrations in nanoscale systems.
%---------------
\begin{figure}
\begin{center}
\includegraphics[width=0.55\textwidth]{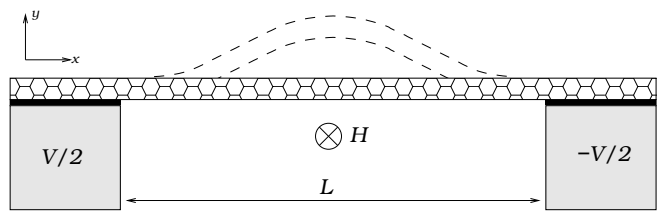}
\end{center}
\caption{Schematic outline of system under consideration. Electronic
  tunnelling through the doubly clamped suspended SWNT under bias
  voltage $V$ excites quantized vibrations of the tube in the presence
  of a magnetic field, $H$. The resulting electromechanical coupling
  leads to an exponential reduction of the probability for tunneling
  through the elastic channel, which, together with Pauli principle
  restrictions on the allowed inelastic transitions, modifies the
  electronic transport through the system at low temperature. The
  result of these effects is both a temperature- and bias
  voltage-independent current deficit (see text) as well as to a
  previously reported negative magnetoconductance
  \cite{Shekhter}. Amplitude shown is greatly exaggerated.}
\label{figure:system}
\end{figure}
%---------------
Also, we find that this reduction of the current is in general not
dependent on strong coupling of the nanowire vibrational modes to the
thermal bath as previously reported in \cite{peapodpaper} where the
current and conductance through a carbon nanotube containing an
encapsulated fullerene was analyzed in the ballistic transport
regime. Hence, the results presented are shown to be quite general
also for oscillator distributions out of equilibrium.

\section{Model}
The system considered is shown \fref{figure:system}, and
comprises a doubly clamped carbon nanotube suspended over a trench of
width $L$, subject to a transverse magnetic field, $H$. In
\cite{Shekhter} it was shown that when the SWNT is biased by a voltage
$V$ the induced mechanical oscillations of the tube lead to
intermediate ``swinging states'' through which electrons can tunnel
between the leads. By restricting the analysis to the fundamental mode
(which gives the most important contribution) the authors showed that
the system can be described by an effective Hamiltonian (Eq. (7) in
\cite{Shekhter}),
%---------------
\begin{equation}
\hat{H}_{eff}=\hat{H}_{leads}+\hat{H}_{osc}+\hat{H}_{T}\,,
\label{Hamiltonian}
\end{equation}
%----------------
which describes charge transfer through the SWNT in the regime of
non-resonant tunnelling or co-tunnelling in the regime of Coulomb
blockade. In \eref{Hamiltonian},
%----------------------
\begin{equation}
\label{Hleads}
\hat{H}_{leads}=\sum_{k;\sigma=l,r}\epsilon_{\sigma,k}\hat{a}^{\dagger}_{\sigma,k}\hat{a}_{\sigma,k}\,,
\end{equation}
%---------------------------
describes the electrons in the leads; $\hat{a}^{\dagger}_{l/r,k}$
[$\hat{a}_{l/r,k}$] are creation [annihilation] operators for
electrons in state $k$ in the left/right lead with energy
$\epsilon_{l/r,k}$ respectively. The second term in the
Hamiltonian, 
%-------------
\begin{equation}
\label{Hosc}
\hat{H}_{osc}=\hbar\omega \hat{b}^{\dagger}\hat{b}\,,
\end{equation}
%-----------------------
describes the oscillating wire where $\hat{b}^{\dagger}$ [$\hat{b}$]
is a boson operator that creates [annihilates] one vibrational quantum
and $\omega=(k/m)^{1/2}$ is the frequency of the fundamental mode of
oscillation with $k$ the rigidity and $m$ the effective mass of the
wire (typically $\omega$ is of the order of~\unit[10$^8$]{s$^{-1}$} if
$L\sim$~\unit[1]{$\mu$m}). The third term in \eref{Hamiltonian}
describes the interaction between the electrons and the oscillating
wire,
%-----------
\begin{equation}
\hat{H}_{T}=e^{i\phi(\hat{b}^{\dagger}+\hat{b})}\sum_{k,k'}T_{eff}(k,k')\hat{a}^{\dagger}_{r,k}\hat{a}_{l,k'}+\textrm{h.c.}\,.
\label{Htunn}
\end{equation}
%---------------
Here, $\phi=4 g\pi x_0 L H/\Phi_0$ is the dimensionless
electron-vibron coupling strength, $\Phi_0=h/e$ is the flux quantum,
$g$ is a geometric factor of order unity and $x_0$ is the zero-point
oscillation amplitude. Finally, $T_{eff}(k,k')$ describes the coupling
of the electronic states $k$ and $k'$ in the different leads due to
tunnelling through virtual states on the wire at zero magnetic
field. The latter are supposed to be discrete due to space
quantization of the electronic motion and possible Coulomb blockade
energy quantization~\footnote{We note that on-wire Coulomb
  interactions only result in a renormalization of the amplitude of
  single electron tunnelling in (and out) of the wire and in
  quantization of its electrostatic charging energy.} (see also
\cite{Shekhter}).

\section{Current}
To calculate the charge transport through the system for the case when
the density matrix is not in thermal equilibrium, we first consider
the time rate of change of the total density matrix for the system,
$\hat{\sigma}(t)$, which is given by the Liouville-von Neumann
equation (see \cite{Fedorets,Novotny2003} for a similar analysis). To
evaluate this we switch to the interaction picture with respect to the
non-interacting Hamiltonian,
$\hat{H}_0=\hat{H}_{leads}+\hat{H}_{osc}$, for which the evolution of
the density matrix is given by $i\hbar\partial_t
\hat{\widetilde{\sigma}}(t)=[\hat{\widetilde{H}}_T(t),\hat{\widetilde{\sigma}}]$,
where $\hat{\widetilde{A}}(t)=e^{i \hat{H}_0 t/\hbar}\hat{A} e^{-i
  \hat{H}_o t/\hbar}$ is any operator in the interaction
picture. Since we are interested in the energy exchange between the
electrons and the oscillating wire we only need to know the evolution
of the reduced density matrix, which is found by tracing out the
degrees of freedom of the leads,
$\hat{\rho}(t)=\Tr_{leads}(\hat{\sigma}(t))$. Treating the electrons
in the leads as fermionic baths whose equilibrium distributions are
virtually unaffected by the charge transfer we evaluate the evolution
of the density matrix to lowest order in the tunnelling
probability. This enables us to find the equation of motion for the
reduced density matrix in the Heisenberg picture,
%--------
\begin{equation}
\label{densityeqn3}
\eqalign{\partial_t \hat{\rho}(t)=&-\frac{i}{\hbar}\left[\hat{H}_{osc},\hat{\rho}(t)\right]\cr
&-\frac{1}{\hbar^2}\Tr\left\{\int_{-\infty}^t\rmd t_1\left[\hat{H}_T,\left[\hat{\widetilde{H}}_T(t_1-t),\hat{\widetilde{\sigma}}(t_1-t)\right]\right]\right\}\,.}
\end{equation}
%--------------------------------------
From this we derive the stationary equation for the reduced
density of the system,
%-----------
\begin{equation}\label{finaldens}
\eqalign{\frac{i}{\hbar}[\hat{H}_{osc},\hat{\rho}]=\vert T_{eff}\vert^2&\Tr\bigg[(\hat{J}_1+\hat{J}_2)\hat{\rho}+\hat{\rho}(\hat{J}_1^{\dagger}+\hat{J}_2^{\dagger})\cr
&- e^{-i \chi \hat{x}}(\hat{J}_1\hat{\rho}+\hat{\rho} \hat{J}_1^{\dagger})e^{i \chi \hat{x}}-e^{i \chi \hat{x}}(\hat{J}_2\hat{\rho}+\hat{\rho} \hat{J}_2^{\dagger})e^{-i \chi \hat{x}}\bigg]\,,}
\end{equation}
%---------------
under the assumption made in \cite{Shekhter} that the overlap integral
$T_{eff}(k,k')$ is independent of the momenta $k$ and $k'$.  In
\eref{finaldens}, $\chi=\sqrt{2}\phi/x_0$, $\hat{x}$ is the deflection
operator of the oscillating wire and the operators $\hat{J}_i$ take on
the form below,
%---------
\begin{eqnarray}
\label{Jfactors}
\fl \hat{J}_{1,2}=-\frac{\nu^2}{\hbar}\int\int\rmd\epsilon_r\rmd\epsilon_l\int_{-\infty}^0\rmd\tau e^{\pm i\chi\hat{x}}e^{i\hat{H}_0\tau}e^{\mp i\chi\hat{x}}e^{-i\hat{H}_0\tau}
f_{r,l}(\epsilon_{r,l})(1-f_{l,r}(\epsilon_{l,r}))e^{\pm i\tau(\epsilon_l-\epsilon_r)}\,.
\end{eqnarray}
%------------
Here, $\nu$ is the density of states in the leads and
$f_{l,r}(\epsilon_{l,r})$ are the Fermi distribution for electrons in
the left/right lead kept at chemical potential $\mu_{l,r}=\pm e V/2$
respectively. Multiplying \eref{finaldens} by the position and
momentum operator and tracing out the oscillator degree of freedom we
find the following expression for the deflection, $\langle
\hat{x}\rangle$, and momentum, $\langle \hat{p}\rangle$, expectation
values,
%---------
\numparts
\label{expect}
\begin{eqnarray}
\label{expectx}
k\langle \hat{x}\rangle=\vert T_{eff}\vert^2 \hbar\chi \textrm{Tr}\left[\left(\hat{J}_r-\hat{J}_l\right)\hat{\rho}\right]\\
\label{expectp}
\langle \hat{p}\rangle=0\,.
\end{eqnarray}
\endnumparts
%-------------- 
\Eref{expectx} gives the force balance in the stationary regime
between the elastic force on the wire (left hand side) and the force
induced by the charge transfer (right hand side) where
$\hat{J}_{r,l}=\hat{J}_{1,2}+\hat{J}^{\dagger}_{1,2}$. On the other
hand a similar expression can be derived from the definition of the
current operator, $\hat{I}=(ie/ \hbar)[\hat{H},\hat{N}_l]$, where
$\hat{N}_l$ is the number operator in the left lead. From the form of
the total Hamiltonian this can be expressed as $i\hbar
m\omega^2\hat{x}+2i\hbar^2\chi\hat{I}/e=[\hat{H},\hat{p}]=-i\hbar\dot{\hat{p}}$~\footnote{The
  same expression can be found by directly evaluating the current
  operator with equation \eref{densityeqn3} to lowest order in the
  tunneling amplitude.}. Under the trace with the static density
matrix the right hand side of this expression vanishes,
$\langle\dot{\hat{p}}\rangle=0$, and we find that the average
mechanical deflection of the wire is proportional to the total current
$I$ through it.

Using this relationship we can thus evaluate the current from
\eref{expectx}. To do so we divide the operators $\hat{J}_{r,l}$ into
their diagonal and non-diagonal parts (subscripts $d$ and $n$
respectively),
$\Tr(\hat{J}_r\hat{\rho})=\Tr(\hat{J}_{r,d}\hat{\rho}_d)+\Tr(\hat{J}_{r,n}\hat{\rho}_n)$,
with respect to the eigenstates of the oscillating wire. From this
analysis we find that to the zeroth order in the operators $\hat{x}$
and $\hat{p}$ the operators $\hat{J}_{r,l}$ only have diagonal
components and the expression for the force is proportional to the
current, $I_0$, through the system (equation (8) in
\cite{Shekhter}). The higher order terms in $\hat{x}$ and $\hat{p}$,
corresponding to the non-diagonal parts of $\hat{J}_{r,l}$, are
collected in the current, $I_1$, which, in the high bias limit, gives
exponentially small corrections to total current $I$ (see below).
%----------------
\begin{equation}
I=I_0+I_{1}\label{current}
\end{equation}
\begin{equation}
\eqalign{I_0=&\frac{G_0}{e}\sum_{n=0}^{\infty}\sum_{\ell=-n}^{\infty}P(n)\vert\langle n\vert e^{i \phi (\hat{b}^{\dagger}+\hat{b})}\vert n+\ell\rangle\vert^2\cr
&\times\int\rmd\epsilon[f_l(\epsilon)(1-f_r(\epsilon-\ell \hbar \omega))-f_r(\epsilon)(1-f_l(\epsilon-\ell \hbar \omega))]\label{unperturbedcurrent}}
\end{equation}
\begin{equation}
I_1=\frac{\vert T_{eff}\vert^2 e}{2}\textrm{Tr}\left[\left(\hat{J}_{r,n}-\hat{J}_{l,n}\right)\hat{\rho}_n\right]\label{currentcorr}\,.
\end{equation}
%----------------------
In \eref{unperturbedcurrent}, $G_0=2e^2 \vert
T_{eff}\vert^2\nu^2/\hbar$ is the zero field conductance and $P(n)$ is
the probability that the fundamental mode is in quantum state $\vert
n\rangle$ with energy $n\hbar\omega$.

The two terms that make up the total current $I$ can be understood as
follows. The first term, $I_0$, is the tunneling current between the
leads which takes into account the coupling between the electronic and
mechanical degrees of freedom. If the distribution of the vibrational
modes are in thermal equilibrium this term is the only contribution to
the current. The second term, $I_1$, on the other hand corresponds to
the back-action on the system due to the electromechanical coupling
and acts to drive the energy distribution of the vibrational modes out
of thermal equilibrium. Thus, for non-zero $I_1$, the distribution
function $P(n)$ in \eref{unperturbedcurrent} is in general not give by
the thermal distribution.

\subsection{Current deficit}
We start the analysis of the current by first considering $I_0$. This
term describes how the combinations of the Fermi functions in the two
leads put restrictions (through the Pauli principle) on the allowed
transmission channels for electrons as they exchange energy with the
vibrating wire. Integrating over the electronic energy this equation
can be expressed as,
%-----------
\begin{equation}
\label{current2}
\fl I_0=\frac{G_0}{e}\sum_{n=0}^{\infty}\sum_{\ell=-n}^{\infty}P(n)\vert\langle n\vert e^{i \phi (\hat{b}^{\dagger}+\hat{b})}\vert n+\ell\rangle\vert^2\left(\frac{\ell \hbar \omega-e V}{e^{\beta(\ell \hbar \omega-e V)}-1}-\frac{\ell \hbar \omega+e V}{e^{\beta(\ell \hbar \omega+e V)}-1}\right)\,.
\end{equation}
%---------------
Here we note that similar expressions for the current have also been
reported for other nanoelectromechanical systems with strong
electron-vibron coupling (e.g., the phenomenon of Franck-Condon
blockade of tunneling through molecular devices
\cite{Koch2005,Braig2003}).

Convergence of the summation over $\ell$ in \eref{current2} is due to
the exponential decay of the matrix element $\langle n\vert
\exp[i\phi(\hat{b}^{\dagger}+\hat{b})]\vert n+\ell\rangle$ at large
$\ell$. Analysis shows that the average number of inelastic scattering
channels, $\bar{\ell}$, that need to be considered in this summation
scales as $\bar{\ell}\propto \phi(\bar{E}/\hbar\omega)^{1/2}$ where
$\bar{E}$ is the average energy associated with the distribution
$P(n)$. This implies that under the condition $V\gg V_0$ ($eV_0\propto
\textrm{max}\{k_bT,\hbar\omega\bar{\ell}\}$) one can neglect the
$\ell$-dependence in the factors $\exp[\beta(\ell\hbar\omega\pm eV)]$
for all relevant $\ell$~\footnote{For self-consistency we also
  evaluate the scaling of the average energy, see \eref{expectenergy}
  below.}. The expression for the current is therefore greatly
simplified in the high bias limit and the sum reduces to
$\sum_{n}\sum_{\ell}P(n)\vert \langle n\vert
e^{i\phi(\hat{b}^{\dagger}+\hat{b})}\vert
n+\ell\rangle\vert^2\ell$. Using the completeness of a set of vibron
states this expression can be directly evaluated to yield,
%--------------------------
\begin{equation}
\eqalign{\fl\sum_{n=0}^{\infty}\sum_{\ell=-n}^{\infty}P(n)\vert \langle n\vert e^{i\phi(\hat{b}^{\dagger}+\hat{b})}\vert n+\ell\rangle\vert^2\ell&=\sum_{n=0}^{\infty}\sum_{n'=0}^{\infty}P(n)\bigg(\langle n\vert e^{i\phi(\hat{b}^{\dagger}+\hat{b})}\hat{b}^{\dagger}\hat{b}\vert n'\rangle\cr
&\qquad\qquad\times\langle n'\vert e^{-i\phi(\hat{b}^{\dagger}+\hat{b})}\vert n\rangle-\langle n\vert \hat{b}^{\dagger}\hat{b}\vert n\rangle\bigg)=\phi^2\,,}
\end{equation}
%--------------------------
which holds for any normalized distribution function $P(n)$. 

The analysis above gives a current deficit through the system at high
bias voltages as compared to the current at zero magnetic
field. Furthermore, the current deficit is found to be independent of
both temperature and the bias voltage.  To understand this one needs
to consider the Pauli restrictions on the inelastic tunneling channels
imposed through equation \eref{unperturbedcurrent}. From this
expression one finds that for low energy electrons many of the
inelastic channels, which act to compensate for the suppression of the
elastic channel, are forbidden. Consequently, the current at low
voltages is reduced from the non-vibrating current, $I=G_0V$, by an
amount that is given by the extent to which the elastic channel is
suppressed. As the voltage increases, more inelastic tunneling
channels are opened and the differential conductance increases
accordingly. In the high voltage limit, $V\gg V_0$, a further increase
of the bias voltage will not be affected by the Pauli restrictions due
to the large energy scales of the electrons, in which case the
differential conductance follows that of the system at zero magnetic
field. Alternatively, this can be viewed as a voltage offset which
depends only on the magnetic field strength and the system's
mechanical parameters,
%-----------
\begin{equation}
\eqalign{I_0(H,V)=G_0\left(V-\frac{\hbar\omega}{e}\phi^2\right)=I_0(0,V-\Delta V(H))\qquad V\gg V_0\cr
\Delta V(H)=16 g^2\pi^2\frac{L^2H^2 e}{m}\,.}
\label{currentdeficit}
\end{equation}
%----------------------
{
%-----------------
\begin{figure}
\begin{center}
\includegraphics[width=0.6\textwidth]{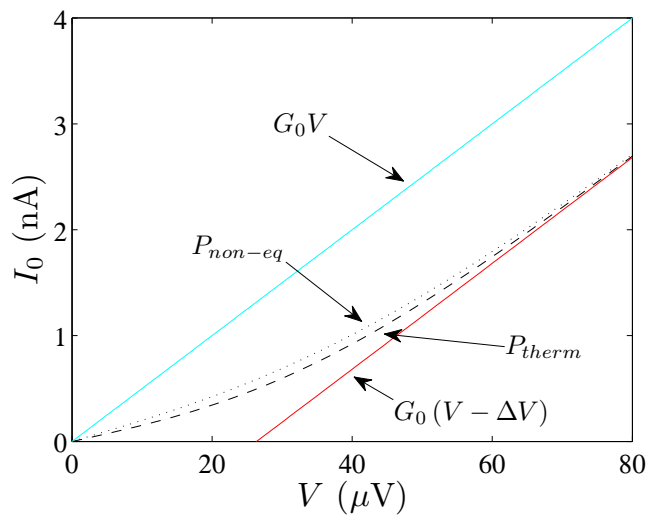} 
\caption{(Colour online) Current as a function of bias voltage
  calculated for two distribution function, $P_{therm}$ (dashed)
  corresponding to the thermal distribution, $P(n)\propto
  e^{-\beta\hbar\omega n}$, and the non-equilibrium distribution
  $P_{non-eq}$ (dotted) where $P(n)\propto
  (eV/\hbar\omega)/((eV/\hbar\omega)^2+n^2)$. Note that $P_{non-eq}$
  may not correspond to the real distribution function but is given as
  an example to reflect the results derived for $\langle
  \hat{p}^2\rangle$ (see text). The current deficit is the difference
  between the current at zero magnetic field, $G_0V$, and the dashed
  and dotted curves respectively (note that the parameters,
  $\omega=$~\unit[10$^{10}$]{s$^{-1}$}, $\phi=2$ and
  $T=$~\unit[0.1]{K} where chosen to clearly separate these two
  curves). Also displayed is the constant voltage offset curve,
  $G_0(V-\Delta V)$.}
\label{fig:current}
\end{center}
\end{figure}
%--------------

\subsection{Higher order corrections to current}
To evaluate the non-diagonal contribution to the current we expand the
exponentials in \eref{Jfactors} in powers of $\hat{H}_0\tau$ and
integrate over the electronic energies,
%----------------------
\begin{equation}
\eqalign{\hat{J}_{1,2}=-\frac{\pi^2 \nu^2}{\hbar}\int_{-\infty}^{0}&\rmd\tau \left(\delta(\tau)+\frac{i}{\beta \sinh(\pi \tau/\beta)}\right)^2\cr
&\times e^{\pm i e V\tau}\sum_{q=0}^{\infty}\left(i \tau\right)^q\frac{1}{q!} e^{\pm i\chi \hat{x}}\hat{Y}(\hat{H}_0,e^{\mp i\chi \hat{x}},q)\,.}
\label{Jfactors2}
\end{equation}
%------------------
In \eref{Jfactors2}, $\hat{Y}(\hat{H}_0,e^{\mp i\chi \hat{x}},q)\equiv
[\hat{H}_0,[\hat{H}_0,[\ldots,e^{\mp i \chi \hat{x}}]]]$ with $q$
indicating the number of commutators to be evaluated and $\beta=(k_B
T)^{-1}$. Evaluating \eref{currentcorr} with this expansion we find
that all contributions to the current $I_1$ decay exponentially in the
high bias limit as all correction terms will be of
the form,
%-------------
\begin{equation}
\eqalign{I_1\propto\int_{-\infty}^{0}\rmd\tau \frac{\tau^q}{\sinh^2(\pi \tau/\beta)}&\left(\begin{array}{cc} \sin(eV\tau)\qquad q=\textrm{odd}\\ \cos(e V\tau)\qquad q=\textrm{even}\end{array}\right)\cr
&\propto (e^{\beta e V}-1)^{-1} \qquad \beta eV\gg 1,q\geq 2\,.}
\end{equation}
%---------------

Thus we find that in the limit of high bias voltage, the current goes
as \eref{currentdeficit}, which differs from the ohmic behaviour for the
non-vibrating wire, $G_0V$, by an amount that is independent of both
the bias voltage and temperature as shown in
\fref{fig:current}. This can be understood from the fact that
the increase in the current due to a further increment in the bias
voltage under the conditions when $V\gg V_0$ is fulfilled is not
affected by the Pauli restrictions on the electron-vibron energy
exchange. Nevertheless, the current deficit at large voltage biases is
a true quantum-mechanical effect on transport that originates from the
Pauli restrictions, however, these restrictions only affect the
tunnelling probability of low energy electrons close to the Fermi
level.

In contrast to the analysis of \cite{peapodpaper} where the
distribution function $P(n)$ was considered to be only slightly out of
equilibrium, our results for the current deficit survives
independently of the form of $P(n)$, even for highly excited
distributions. As an example, we have analyzed \eref{finaldens}
and \eref{Jfactors} separately to order $\hat{p}$, an analysis which
shows that in the high bias limit the distribution function is indeed
far from equilibrium as, e.g., the magnitude of the two lowest
non-zero moments are~\footnote{The coefficients multiplying higher
  order terms in this expansion are exponentially small in the limit
  of high bias voltage and are thus ignored.},
%------------
\numparts
\label{expectscaling}
\begin{eqnarray}
\beta e V\gg 1\nonumber\\
\langle \hat{x}\rangle\sim \frac{\sqrt{2} \pi G_0 \phi x_0}{e \omega} V\\
\label{expectenergy}
\langle \hat{p}^2\rangle\sim \frac{\hbar}{4 x_0^2 \omega} (e V-2 \hbar\omega \phi^2)\,.
\end{eqnarray}
\endnumparts
%------------

%-----------------
\begin{figure}
\begin{center}
\includegraphics[width=0.6\textwidth]{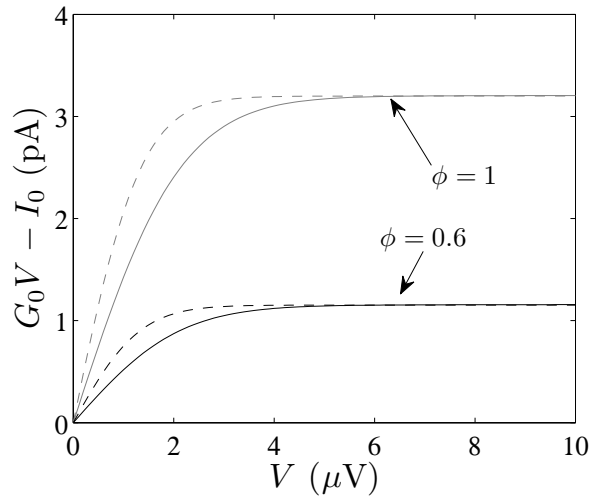}
\caption{Plot of current deficit as a function of bias voltage. Here,
  $\omega=$ \unit[10$^8$]{s$^{-1}$}, $\phi=1$ (gray), $\phi=0.6$
  (black), $T=$ \unit[8]{mK} (solid) and $T=$ \unit[5]{mK} (dashed).}
\label{fig:currentreduction}
\end{center}
\end{figure}
%--------------
Finally, we show the current deficit as a function of the bias voltage
for realistic experimental parameters,
\fref{fig:currentreduction}. Here, the influence of the
multiconnectivity of the electron tunnelling paths in the elastic
channel and the Pauli-principle restrictions on the available
inelastic channels are clearly visible at low bias voltages as an
increasing current deficit. Eventually, when the new inelastic
tunnelling channels that are added by a further increment of the bias
voltage are not affected by the Pauli restrictions --- which occurs at
large enough bias voltages --- the current deficit saturates to a
constant value. This constant value of the current deficit depends on
the magnetic field $H$ and the mechanical properties of the nanowire
through the parameter $\phi$ as illustrated in
\fref{fig:currentreduction}.

\section{Conclusion}
The analysis followed here assumes non-resonant tunnelling or
co-tunnelling (for the case of strong Coulomb interactions) of
electrons through the wire. For the realistic example of a suspended
carbon nanotube of length $L\sim$~\unit[1]{$\mu$m} we estimate that
this can be achieved for bias voltages below a few~\unit[]{mV}. For
the same system we estimate that the current offset should be of the
order of a few picoampere for magnetic fields of the order of
$H\sim$~\unit[20]{T} and bias voltages $V\sim$~\unit[10]{$\mu$V} as
shown in \fref{fig:currentreduction} (see also discussion in
\cite{Shekhter}). Finally the temperature range necessary for these
effects to be observable requires that $T\ll eV/k_B$ (typically a
few~\unit[]{mK}) to avoid smearing out of the Fermi distributions as
well as back-action from the wire on the electronic system. Should
this not be the case, the main result of this paper does not apply as
the current deficit depends crucially on the extent to which the Pauli
principle puts restrictions on the allowed electronic tunneling
channels.

As discussed above, the role of electron-electron interactions on the
wire can be shown to lead only to an effective renormalization of the
amplitude of single electron tunnelling on (and off) the wire. In
order to verify this we suggest an experimental protocol where one
measures the voltage offset at finite magnetic field,
%--------------------
\begin{equation}
\Delta V(H)=\frac{I_0(0,V)-I_0(H,V)}{G_0}\qquad V\gg V_0\,,
\end{equation}
%-------------------
from which the zero field conductance, $G_0$, can be deduced as
$\Delta V$ is only a function of the system parameters and the
magnetic field strength.

Concluding, we have shown that for the system originally considered by
Shekhter \etal \cite{Shekhter} not only will the conductance be
altered due to the multiconnectivity of the electronic transport
through the system, but also the current. In particular we find that
even for vibrational distributions out of thermal equilibrium the
system displays a current deficit which is independent of bias voltage
and temperature. This is a clear manifestation of quantum mechanical
effects on transport not previously considered. Also, we have shown
that the influence of internal damping in this nonresonant charge
transfer process decays exponentially to all orders in the moments of
the position and momentum, thus making the system considered a very
good candidate for direct observation of quantum mechanical effects in
mesoscopic systems.

\section*{Acknowledgments}
\addcontentsline{toc}{section}{Acknowledgments} Discussions with
Leonid Gorelik, Robert Shekhter, Ilya Krive and Mats Jonson are
gratefully acknowledged. This work was supported in part by the
Swedish VR and SSF.

\section*{References}
\addcontentsline{toc}{section}{References}
\bibliographystyle{unsrt}
\bibliography{/chalmers/users/sonneg/Desktop/Artiklar/referencesAharanov}
%-------------

\end{document}